# A Forecasting-Based DLP Approach for Data Security

**Kishu Gupta and Ashwani Kush**

**Abstract** Sensitive data leakage is the major growing problem being faced by enterprises in this technical era. Data leakage causes severe threats for organization of data safety which badly affects the reputation of organizations. Data leakage is the flow of sensitive data/information from any data holder to an unauthorized destination. Data leak prevention (DLP) is set of techniques that try to alleviate the threats which may hinder data security. DLP unveils guilty user responsible for data leakage and ensures that user without appropriate permission cannot access sensitive data and also provides protection to sensitive data if sensitive data is shared accidentally. In this paper, data leakage prevention (DLP) model is used to restrict/grant data access permission to user, based on the forecast of their access to data. This study provides a DLP solution using data statistical analysis to forecast the data access possibilities of any user in future based on the access to data in the past. The proposed approach makes use of renowned simple piecewise linear function for learning/training to model. The results show that the proposed DLP approach with high level of precision can correctly classify between users even in cases of extreme data access.

## 1 Introduction

The NIST explains computer security as "protection afforded to an automated information system in order to attain the applicable objectives of preserving the integrity, availability, and confidentiality of information system resources (includes hardware,

K. Gupta (✉)
Department of Computer Science and Applications, Kurukshetra University, Kurukshetra, India
e-mail: kishugupta2@gmail.com

A. Kush
Institute of Integrated and Hons. Studies, Kurukshetra University, Kurukshetra, India
e-mail: akush20@gmail.com

software, firmware, information/data, and telecommunications)" [1, 2]. Advancement in technology allows easy and speedy transfer of data. Data is the key to conduct business activities nowadays, and hence, a need arises to share data among various stakeholders/third parties like human resources working from outside the site (e.g., on laptops), business colleague, and clients [3]. For example, service provider requires access to the company intellectual property and other confidential information to carry out their services [4, 5].

Data loss/leakage has emerged as the biggest threat that organizations are facing today. In the present scenario, almost all business activities depend on extensive sharing of sensitive/confidential data, within or outside the organization [6, 7]. Data leakage is an event that may occur either accidentally or maliciously that permits data access to unauthorized user. Sensitive data loss/leakage rigorously hampers reputation of organization, confidence/faith of customers in company which may ultimately lead to shut down company or even may lead to severe political crisis like WikiLeak leaks [8]. Leakage is thus a subset of data loss with a spotlight on the data security goal.

To minimize the risk of data loss, organizations usually make use of DLP solutions as a protection/defense mechanism. Prior to DLP security, mechanisms/technologies like firewalls and IDS were in use [9]. DLPSs are used to protect all kind of data, that is, data in use, data at rest and data in transit. DLPSs use the statistical/ analytical approach, data fingerprinting, regular expressions on context and content of data to identify and avoid unauthorized access to data [10]. DLP approach performs deep content analysis and observes the data access by users to discover improper usage [11–13]. DLP systems employ a model using either knowledge of an expert or may train/learn from available past records (Fig. 1).

This scenario provides ample space to produce requirement of a mechanism that can identify leakage with more precision for greater data security. The proposed DLP model tries to provide data security by observing users' trend to access the date, uses learning-based approach to highlight the user whoever is performing different to trend observed. This enables organization to take suitable action like imposing access restriction on sensitive data for data security.

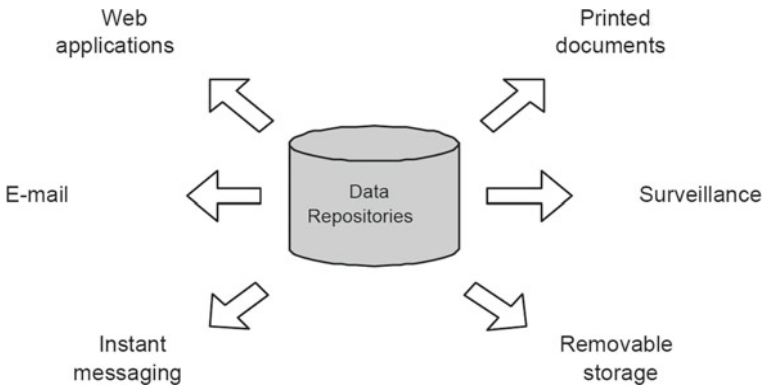

**Fig. 1** Possible leak channels [14]

The paper layout is as follows: in Sect. 2, an overview of DLP solutions related to work has been presented. Furthermore, an overview of the proposed data fitting model framework has been discussed in Sect. 3. Section 4 deals with the experimental results, and Sect. 5 represents conclusion of paper, respectively.

## 2 Overview of DLP Approach

This section specifies major benefits of DLP solution. DLP model generally distinguishes suspicious activity from normal activity and performs either detection, i.e., raise alert if doubtful activity happens, or prevention, i.e., block nasty activity. DLP model can be characterized by various dimensions like model construction, filtering approach, network-based, host-based, etc. Model building approach to describe how the model is constructed is most relevant and best suited for the proposed work. Specification and learning-based are two approaches for model construction. Specification-based approach uses expert's knowledge and hence more precise/accurate, while learning-based model automatically learns using statistical techniques, i.e., machine learning.

The proposed framework reflects numerous benefits over existing solutions for DLP. First, learning-based framework tailors itself to the user's behavior and hence makes feasible to detect unknown and insider attacks. Additionally proposed DLP approach provides better control on data from being misused along with providing flexibility to access the data simultaneously. Moreover, this approach integrates data protection with user identity, thus making organization capable to implement data protection policy based on user identity and their role. The proposed approach tries to forecast guilty user based on available records of user access to the organization data. Finally, this forecasting-based framework proves to be more practical and efficient (Fig. 2).

## 3 Data Training Model

The proposed DLP model fulfills the objective of data security by employing machine learning-based approach and provides forecast for further action. The proposed model considers multiple agents (m) which have accessibility to organization data anytime and for any number of times. Each and every access to organization data by any user is entered in the form of a user's accessibility dataset containing many important details like date to access data, duration for which organization data was in use by particular user, i.e., *y*. The user's accessibility dataset continues to grow with time and is fitted and trained by machine learning techniques to obtain trend. Trend in this study determines the pattern of data accessibility by various users along a period of time. Thus, the proposed model provides future insight by studying the past events. This section uses dataset of 2014 to 2018 to train the model and then predicts the trend of particular user after 2018. If the trend exceeds the defined upper limit, then it raises alert to prevent or restrict the user access to organization data. The discussed approach is explained through Fig. 3.

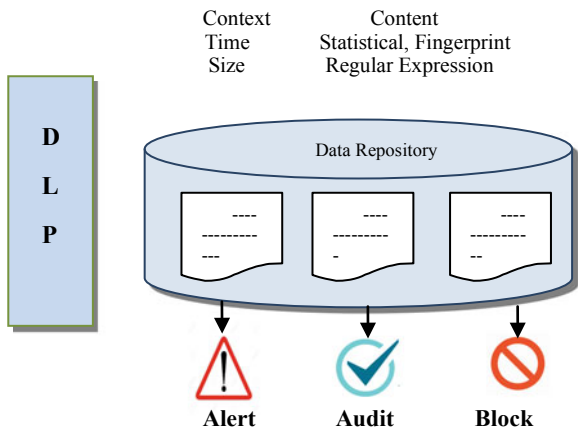

**Fig. 2** DLP overview

### 3.1 Model Equations

The model uses a simple piecewise linear model-based function as shown in Eq. (1) named as model fitting equation.

$$\mathscr{g}(t) = (k + \alpha(t)'\delta)t + (m + \alpha(t)'\gamma). \tag{1}$$

The above model fitting equation is used to fit and train the user's accessibility dataset to evaluate the predicted time value to be spent by user $\hat{y}$ along with many other parameters required for study. The model executes all entries in user accessibility database, i.e., $i = 1$ to $n$.

### 3.2 Model Accuracy

After obtaining predicted time value, the model calculates the error existing between actual and predicted value to determine the accuracy of approach. Here, $\epsilon_i$ shown in Eq. (2A) is error existing between actual and predicted values of time spend by user to access data. $£_i$ is percentage error represented by Eq. (2B).

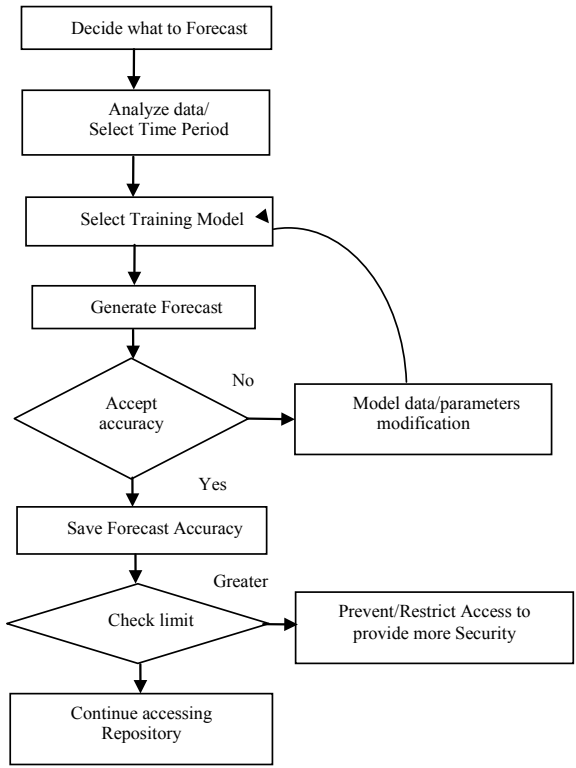

Fig. 3 Data training model

$$\epsilon_i = \left((y_i + \alpha) - (\hat{y}_i - \alpha)\right) \tag{2A}$$

$$\epsilon_i = 100 * \frac{\epsilon_i}{(y_i + \alpha)}. \tag{2B}$$

## 3.3 Calculating Limit

To define the upper bound and lower bound for a user to access the data, the study computes $Æ_i$, $L_i$ and $υ_i$ in Eqs. (3A, 3B and 3C), respectively.

$$Æi = |\varepsilon i|. \tag{3A}$$

$$Li = \left(\hat{y}i - (\mu * \varsigma * \sigma)\right). \tag{3B}$$

$$Ui = \left(\hat{y}i + (\mu * \varsigma * \sigma)\right). \tag{3C}$$

Æ$_i$ is absolute error, i.e., nonnegative value of error as calculated in Eq. (2A). $L_i$ is lower bound and $υ_i$ is upper bound for users to access data. Here, м and $σ$ are mean and standard deviation of Æ calculated as shown below.

$$ μ = \frac{\sum_{i=1}^{n} Æi}{n} \text{ and } σ = \sqrt{\frac{1}{n}\sum_{i=1}^{n}(Æi - \overline{Æ})^2} $$

## 4 Experimental Result

The proposed model is implemented using Python on Jupyter Notebook platform in Anaconda environment to conduct the experiments. Figure 4a shows the overall half yearly forecast, and Fig. 4b shows overall annual forecast; it is observed that from 2014 to 2018, data was accessed for less than 80 min, given a year or half yearly. These graphs show access time forecast of overall system, i.e., all users together. Separate graphs for particular users can also be generated. Forecasting results depict that some user is going to access the data for longer duration in 2019 which is unusual as compared to the previous years. Based on trend, it can be a scenario/possibility of data leakage; hence, more restriction can be imposed to the user by checking on the access time limit for particular user. This is how this model will help to prevent data leakage for database being in use by multiple users and multiple repositories.

## 5 Conclusion

The biggest challenge in the present era is to shield sensitive data from leakage, which imposes a big threat for organization's growth/security/health. The paper highlights DLP approach with better control on data to protect data from being misused and also provide flexibility to access the data simultaneously. Moreover, this approach integrates data protection with user identity, hence enabling organization to enforce data protection policy based on user identity and their role. The proposed approach tries to forecast seems to be guilty user based on available records of user access to the organization data. On basis of model prediction output; access rights of any particular user can be restricted or blocked completely, hence proposed model provides enhanced data security for organization data. Thus, the proposed model provides future insight by studying the past events. Conclusion drawn from this study is that the system based on forecasting approach to identify guilty user is more practical and efficient.

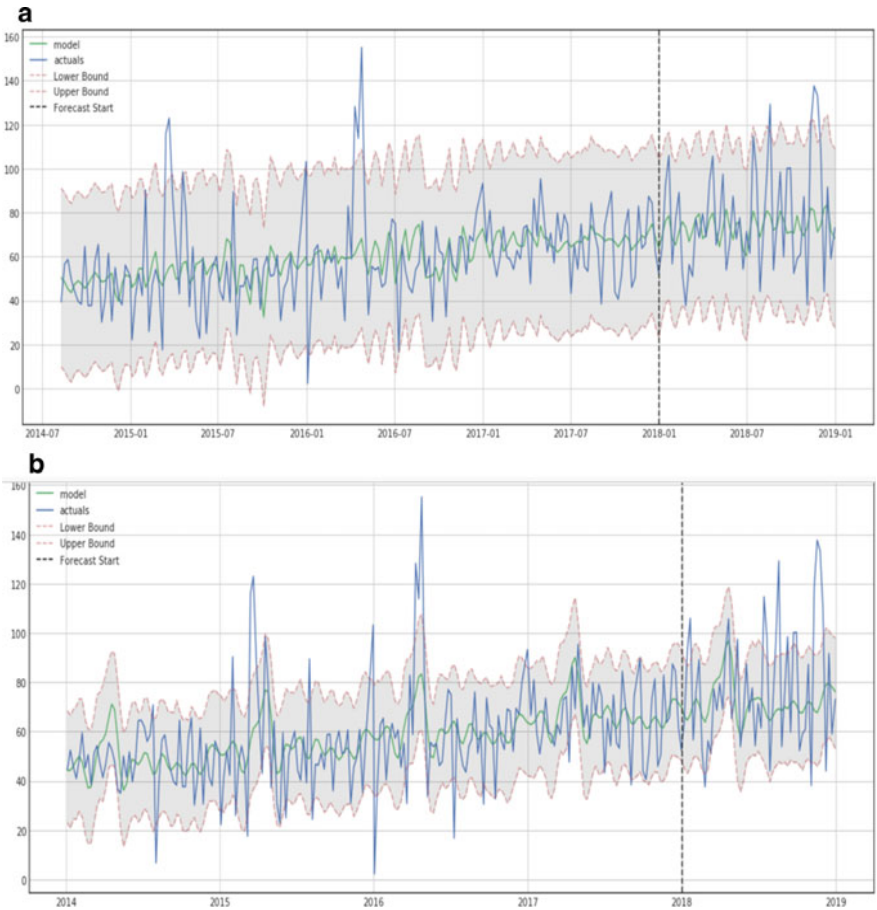

**Fig. 4** **a** Overall half yearly forecast, **b** overall annual forecast